\documentclass[12pt]{article}
\begin{document}
\begin{center}
{\large\bf Noncommutative and Non-Anticommutative Quantum Field Theory}
\vskip 0.3 true in
{\large J. W. Moffat}
\vskip 0.3 true in
{\it Department of Physics, University of Toronto,
Toronto, Ontario M5S 1A7, Canada}
\end{center}
\begin{abstract}%
A noncommutative and non-anticommutative quantum field theory is formulated
in a superspace, in which the superspace coordinates satisfy noncommutative
and non-anticommutative relations. A perturbative scalar field theory is
investigated in which only the non-anticommutative algebraic structure is
kept, and one loop diagrams are calculated and found to be finite due to
the damping caused by a Gaussian factor in the propagator.
\end{abstract}



\section{Introduction}

Noncommutative geometry is characterized by $d$ noncommuting
self-adjoint operators ${\hat x}^\mu$ in a Hilbert space ${\cal H}$
satisfying~\cite{Connes,Witten}
\begin{equation}
[{\hat x}^\mu,{\hat x}^\nu]=i\theta^{\mu\nu},
\end{equation}
where
$\theta^{\mu\nu}$ is a non-degenerate $d\times d$ antisymmetric matrix. The
product of two operators ${\hat\phi}_1$ and ${\hat\phi}_2$ has a
corresponding Moyal $\star$-product
\begin{equation}
({\hat\phi}_1\star{\hat\phi}_2)(x)
=\biggl[\exp\biggr(\frac{i}{2}\theta^{\mu\nu}\frac{\partial}
{\partial\xi^\mu}\frac{\partial}
{\partial\zeta^\nu}\biggr)\phi_1(x+\xi)\phi_2(x+\zeta)\biggr]_{\xi=\zeta=0}.
\end{equation}

Recently, the consequences for gravitation theory of using the Moyal
$\star$-product to define a general quantum gravity theory on a
noncommutative spacetime were analyzed, and it was found that a complex
symmetric metric (non-Hermitian) theory could possibly provide a consistent
underlying geometry on a complex coordinate manifold~\cite{Moffat}.
In a subsequent paper~\cite{Moffat2} the gravitational
Einstein-Hilbert action given on a noncommutative spacetime geometry was
expanded about a flat Minkowski spacetime, and the products of the fields
were replaced by Moyal $\star$-products.
It was found that the planar one loop graviton graph and vacuum
polarization are essentially the same as for the commutative
perturbative result, while the non-planar graviton loop graph is
damped due to the oscillatory behavior of the noncommutative
phase factor in the Feynman integrand. Thus, the overall
noncommutative perturbative quantum gravity theory remains
unrenormalizable and divergent. The free-field theory
in noncommutative field theories is the same as the commutative
field theories, so only the interactions contain new
information. Thus, all the planar graphs at all orders of
perturbation theory are the same as in the commutative theory,
except for unimportant vertex phase factors that depend only on
the external momenta. There is no non-trivial dependence on
internal momenta phase factors, which could help to dampen the
divergent behavior of planar diagram loop integrals.

In the following, we shall investigate a spacetime
geometry in which the spacetime coordinates are
treated as `superspace' vector operators in a superspace with
`fermionic' degrees of freedom, as well as the familiar `bosonic',
commuting or noncommuting degrees of freedom. A
calculation of the one loop diagrams is carried out in scalar field theory.
The dynamics of Yang-Mills gauge field theories and perturbative quantum
gravity will be studied in a future publication.

\section{Superspace Formalism}

We shall formulate our quantization of spacetime on a superspace
~\cite{DeWitt,Buchbinder} based on an extended Grassmann
algebra ${\cal A}$. Let us define the superspace coordinates
\begin{equation}
\label{supercoordinates}
\rho^\mu=x^\mu+\beta^\mu,
\end{equation}
where $x^\mu$ denote the familiar spacetime coordinates saisfying
\begin{equation}
[x^\mu,x^\nu]=0,
\end{equation}
and $\beta^\mu$ are coordinates in an associative Grassman
algebra which satisfy
\begin{equation}
\{\beta^\mu,\beta^\nu\}\equiv\beta^\mu\beta^\nu+\beta^\nu\beta^\mu=0.
\end{equation}
Here, $\mu=0...3$ although the formalism can easily be
extended to higher-dimensional spaces.

The product of two operators ${\hat\phi}_1$ and ${\hat\phi}_2$
in our superspace is given by the $\circ$-product
\begin{equation}
\label{phiproduct}
({\hat\phi}_1\circ{\hat\phi}_2)(\rho)=\biggl[\exp\biggl(\frac{1}{2}\omega^{\mu\nu}
\frac{\partial}{\partial\rho^\mu}\frac{\partial}{\partial\eta^\nu}\biggr)
\phi_1(\rho)\phi_2(\eta)\biggr]_{\rho=\eta} $$ $$
=\phi_1(\rho)\phi_2(\rho)+\frac{1}{2}\omega^{\mu\nu}\frac{\partial}{\partial\rho^\mu}
\phi_1(\rho)\frac{\partial}{\partial\rho^\nu}\phi_2(\rho)+O(\omega^2),
\end{equation}
where $\omega^{\mu\nu}$ is a nonsymmetric tensor
\begin{equation} \omega^{\mu\nu}=-\tau^{\mu\nu}+i\theta^{\mu\nu},
\end{equation}
with $\tau^{\mu\nu}=\tau^{\nu\mu}$ and
$\theta^{\mu\nu}=-\theta^{\nu\mu}$. Moreover, $\omega^{\mu\nu}$ is
Hermitian symmetric $\omega^{\mu\nu}=\omega^{\dagger\mu\nu}$, where
$\dagger$ denotes Hermitian conjugation.

Let us define the notation
\begin{equation}
[\phi_1(\rho),\phi_2(\rho)]_\circ\equiv
\phi_1(\rho)\circ\phi_2(\rho)-\phi_2(\rho)\circ\phi_1(\rho),
\end{equation}
and
\begin{equation}
\{\phi_1(\rho),\phi_2(\rho)\}_\circ\equiv
\phi_1(\rho)\circ\phi_2(\rho)+\phi_2(\rho)\circ\phi_1(\rho).
\end{equation}
From (\ref{supercoordinates}) and (\ref{phiproduct}), we obtain for the
superspace operator ${\hat\rho}$:
\begin{equation}
\label{circnoncom}
[{\hat\rho}^\mu,{\hat\rho}^\nu]_\circ=2\beta^\mu\beta^\nu+i\theta^{\mu\nu}+O(\theta^2),
\end{equation}
\begin{equation} \label{circnonanticom}
\{{\hat\rho}^\mu,{\hat\rho}^\nu\}_\circ=2x^\mu
x^\nu+2(x^\mu\beta^\nu+x^\nu\beta^\mu) -\tau^{\mu\nu}+O(\tau^2).
\end{equation}

If we now choose the limit in which $\beta^\mu\rightarrow 0$ and
$\tau^{\mu\nu}\rightarrow 0$, then we get from (\ref{circnoncom}) and
(\ref{circnonanticom}):
\begin{equation}
\label{standardnoncom}
[{\hat\rho}^\mu,{\hat\rho}^\nu]_\circ\rightarrow [{\hat x}^\mu,{\hat
x}^\nu]=i\theta^{\mu\nu},
\end{equation}
\begin{equation}
\label{standardanti}
\{{\hat\rho}^\mu,{\hat\rho}^\nu\}_\circ\rightarrow \{{\hat x}^\mu,{\hat
x}^\nu\}=2x^\mu x^\nu.
\end{equation}
We see that (\ref{standardnoncom}) and (\ref{standardanti})
give us back the usual noncommutative expressions for the coordinate
operators ${\hat x}^\mu$.

In the limit $x^\mu\rightarrow 0$ and $\theta^{\mu\nu}\rightarrow 0$, we get from
(\ref{circnoncom}) and (\ref{circnonanticom}):
\begin{equation}
[{\hat\rho}^\mu,{\hat\rho}^\nu]_\circ\rightarrow
[{\hat\beta}^\mu,{\hat\beta}^\nu]=2\beta^\mu\beta^\nu,
\end{equation}
\begin{equation}
\label{rhoequation}
\{{\hat\rho}^\mu,{\hat\rho}^\nu\}_\circ\rightarrow
\{{\hat\beta}^\mu,{\hat\beta}^\nu\}=-\tau^{\mu\nu}.
\end{equation}
Let us now perform
a similarity transformation on ${\hat\beta}^\mu$ to the Dirac $\gamma^\mu$ matrices,
and choose the canonical diagonalized form $\tau^{\mu\nu}=2\eta^{\mu\nu}$. We now
obtain from (\ref{rhoequation}): \begin{equation}
\{{\hat\beta}^\mu,{\hat\beta}^\nu\}=\{\gamma^\mu,\gamma^\nu\} =-2\eta^{\mu\nu},
\end{equation} where the $\gamma^\mu$ form a Clifford algebra ${\cal A}_\gamma$ over
the real numbers ${\cal R}$~\cite{Tyutin,Gaetano}.

In the following, we shall consider the simpler geometry determined by
$\theta^{\mu\nu}=0$. We define a $\diamondsuit$-product
\begin{equation}
\label{diamondproduct}
({\hat\phi}_1\diamondsuit{\hat\phi}_2)(\rho)=\biggl[\exp\biggl(-\frac{1}{2}\tau^{\mu\nu}
\frac{\partial}{\partial\rho^\mu}\frac{\partial}{\partial\eta^\nu}\biggr)
\phi_1(\rho)\phi_2(\eta)\biggr]_{\rho=\eta} $$ $$
=\phi_1(\rho)\phi_2(\rho)-\frac{1}{2}\tau^{\mu\nu}\frac{\partial}{\partial\rho^\mu}
\phi_1(\rho)\frac{\partial}{\partial\rho^\nu}\phi_2(\rho)+O(\tau^2).
\end{equation}
We now have
\begin{equation}
\{{\hat{\rho}^\mu,{\hat\rho}^\nu}\}=2x^\mu
x^\nu+2(x^\mu\beta^\nu+x^\nu\beta^\mu)-\tau^{\mu\nu}.
\end{equation}

\section{\bf Non-Anticommutative Field Theory}

Our superspace formulation is a unified description of both noncommutative
and non-anticommutative operator algebras in a Hilbert space ${\cal H}$.
We could formulate a general field theory using the $\circ$-product of
fields, which would contain complex exponential phase factors from the
contributions associated with the antisymmetric tensor $\theta^{\mu\nu}$, as well as the
real exponential factors associated with the contributions from the
symmetric tensor $\tau^{\mu\nu}$. However, in the following, we shall
restrict ourselves to the simpler case when $\theta^{\mu\nu}=0$.

For the $\diamondsuit$-product of exponential functions, we have the rule
\begin{equation}
\exp(ik\rho)\diamondsuit\exp(iq\rho)=\exp[i(k+q)\rho]\exp[\frac{1}{2}(k\tau
q)],
\end{equation}
where $(k\tau q)\equiv k_\mu\tau^{\mu\nu}q_\nu$. We have for
two operators ${\hat f}$ and ${\hat g}$:
\begin{equation}
({\hat f}\diamondsuit {\hat g})(\rho)=\frac{1}{(2\pi)^8}\int d^4k
d^4q{\tilde f}(k){\tilde g}(q)\exp\biggl[\frac{1}{2}(k\tau
q)\biggr]\exp[i(k+q)\rho],
\end{equation}
where
\begin{equation}
{\tilde f}(k)=\frac{1}{(2\pi)^4}\int d^4\rho\exp(-ik\rho)f(\rho).
\end{equation}
The product rule satisfies the
associative rule
\begin{equation}
(f\diamondsuit g)\diamondsuit
h(\rho)=f\diamondsuit(g\diamondsuit h)(\rho).
\end{equation}
We shall assume that derivatives act trivially in our space:
$\partial/\partial\rho^\mu=\partial_\mu$
and
\begin{equation}
[\rho^\mu,\partial_\nu]=-\eta_{\mu\nu},\quad
[\partial_\mu,\partial_\nu]=0.
\end{equation}

In contrast to the noncommutative field theories, the kinetic energy
component of the action in non-anticommutative field theories is not
trivially the same as the commutative field theories, because of the
symmetry of the tensor $\tau^{\mu\nu}$. The action for the scalar field
$\phi$ is
\begin{equation}
S=\int
d^4\rho\biggl[\frac{1}{2}\partial_\mu\phi(\rho)\diamondsuit\partial^\mu\phi(\rho)
-\frac{m^2}{2}\phi(\rho)\diamondsuit\phi(\rho)-V_\diamondsuit(\phi)\biggr],
\end{equation}
where $V_\diamondsuit(\phi)$ is the scalar field potential.
For our calculations we shall choose the potential: \begin{equation}
\label{potential}
V_\diamondsuit(\phi)=\frac{\lambda}{4!}\phi\diamondsuit\phi\diamondsuit\phi\diamondsuit\phi.
\end{equation}

In the noncommutative and the non-anticommutative cases there is an
ambiguity in applying the quantization procedure in position superspace.
The usual quantization condition
\begin{equation}
[\phi({\vec\rho},t),\pi({\vec \eta},t)]=i\delta^{(3)}({\vec \rho}-{\vec \eta}),
\end{equation}
is defined for $\phi$ and $\pi$ in different superspace
points, while the $\star$ and $\diamondsuit$-products only make sense when
the products are computed at the same superspace point. We can avoid this
problem by working only in momentum space when performing Feynman rule
calculations to obtain matrix elements. With this in mind, we shall
formally define a $\bigtriangleup$-product of field operators ${\hat
f}(\rho)$ and ${\hat g}(\eta)$ at different superspace points
$\rho$ and $\eta$:
\begin{equation}
{\hat f}(\rho)\bigtriangleup {\hat g}(\eta)
=\exp\biggl(-\frac{1}{2}\tau^{\mu\nu}\frac{\partial}{\partial
\rho^\mu}\frac{\partial}{\partial \eta^\nu}\biggl)f(\rho)g(\eta).
\end{equation}
This product reduces to the $\diamondsuit$-product of ${\hat
f}$ and ${\hat g}$ in the limit $\rho\rightarrow \eta$.

For the $\diamondsuit$-product rule in momentum space, we have
the quantization condition
\begin{equation}
[{\tilde\phi}(k),{\tilde\pi}(q)]_\diamondsuit=i\delta^{(3)}(k-q)
\exp[\frac{1}{2}(k\tau q)].
\end{equation}
In the case of the non-anticommutative field
theory the exponential factor is relevant, due to the symmetry of
the tensor $\tau^{\mu\nu}$, whereas in the noncommutative field
theory the corresponding phase factor $\exp[-\frac{i}{2}(k\theta
q)]$ has no relevance because of the antisymmetry of the tensor
$\theta^{\mu\nu}$ and the definition $(k\theta q)\equiv
k_\mu\theta^{\mu\nu} q_\nu$.

Let us now consider the Feynman rules for the non-anticommutative
scalar field theory. We shall begin by calculating the two-point
propagator. We quantize the scalar field in the same manner
as in commutative field theory. We choose the free field operator
$\phi(\rho)$ to have the form
\begin{equation}
\phi(\rho)=\frac{1}{\sqrt{2(2\pi)^3}}\int
\frac{d^3k}{k_0}[a(k)\exp(-ik\rho)+a^\dagger(k)\exp(ik\rho)],
\end{equation}
where the annihilation and creation operators
satisfy
\begin{equation}
[a(k),a(k')]=[a^\dagger(k),a^\dagger(k')]=0,\quad
[a(k),a^\dagger(k')] =\delta^{(3)}({\vec k}-{\vec k}').
\end{equation}
We now find using the $\bigtriangleup$-product
\begin{equation}
[\phi(\rho),\phi(\eta)]_\bigtriangleup
\equiv i{\bar\Delta}(\rho-\eta)=
\exp\biggl(-\frac{1}{2}\tau^{\mu\nu}
\frac{\partial}{\partial \rho^\mu}\frac{\partial}{\partial
\eta^\nu}\biggr)[\phi(\rho),\phi(\eta)] $$ $$
=\frac{1}{(2\pi)^3}\int\frac{d^3k}{k_0}\exp[-ik(\rho-\eta)]\epsilon(k_0)
\exp[\frac{1}{2}(k\tau k)]\delta(k^2-m^2),
\end{equation}
where $\epsilon(k_0)=k_0/\vert
k_0\vert$ is $=+1$ if $k_0>0$ and $-1$ if $k_0 <0$. Also, we have
\begin{equation}
\label{commutator}
[\phi(\rho),\phi(\eta)]_\bigtriangleup=\phi(\rho)\bigtriangleup\phi(\eta)-\phi(\eta)
\bigtriangleup\phi(\rho).
\end{equation}
Moreover, we have
\begin{equation}
\biggl[i{\bar\Delta}(\rho)\biggr]_{\vert\tau^{\mu\nu}\rightarrow 0}
=i\Delta(\rho),
\end{equation}
where $\Delta$ is the familiar commutative
theory Green's function
\begin{equation}
i\Delta(\rho)=\frac{1}{(2\pi)^3}\int\frac{d^3k}{k_0}\exp(-ik\rho)\epsilon(k_0)
\delta(k^2-m^2).
\end{equation}

The modified Feynman propagator ${\bar\Delta}_F$ is defined by
the vacuum expectation value of the time-ordered
$\bigtriangleup$-product
\begin{equation}
i{\bar\Delta}_F(\rho-\eta)\equiv\langle 0\vert
T(\phi(\rho)\bigtriangleup\phi(\eta))\vert 0 \rangle
$$ $$
=\frac{i}{(2\pi)^4}\int\frac{d^4k\exp[-ik(\rho-\eta)]\exp[\frac{1}{2}(k\tau
k)]}{k^2-m^2+i\epsilon}.
\end{equation}
In momentum space this
gives
\begin{equation}
i{\bar\Delta}_F(k)=\frac{i\exp[\frac{1}{2}(k\tau k)]}
{k^2-m^2+i\epsilon},
\end{equation}
which reduces to the standard
commutative field theory form for the Feynman propagator
\begin{equation}
i\Delta_F(k)=\frac{i}{k^2-m^2+i\epsilon}
\end{equation} in
the limit $\vert\tau^{\mu\nu}\vert\rightarrow 0$.
 
The interaction part of the action gives
\begin{equation}
S_{\rm int}=\frac{\lambda}{4!}\int
d^4\rho\phi\diamondsuit\phi\diamondsuit\phi\diamondsuit\phi(\rho)
=\frac{\lambda}{4!}\int d^4k_1...d^4k_4\exp[\frac{1}{2}(k_1\tau
k_2)]\exp[\frac{1}{2}(k_2\tau k_3)]
$$ $$
\exp[\frac{1}{2}(k_3\tau k_4)]\phi(k_1)
\phi(k_2)\phi(k_3)\phi(k_4)
(2\pi)^4\delta^{(4)}(k_1+k_2+k_3+k_4).
\end{equation}
From this result, we can deduce that the vertex
factor for the scalar
non-anticommutative theory is given by
\begin{equation}
V(k_1,k_2,k_3,k_4)=\exp\biggl\{\frac{1}{2}[(k_1\tau k_2)
+(k_2\tau k_3)+(k_3\tau k_4)]\biggr\}.
\end{equation}
The vertex
function factor for a general diagram has the form
\begin{equation}
V(k_1,...,k_n)=\sum_{i,j}\exp\biggl[\frac{1}{2}C_{ij}(k_i\cdot k_j)\biggr],
\end{equation}
where $k_i\cdot k_j=k_{i\mu}\tau^{\mu\nu}k_{j\nu}$
and $C_{ij}$ is a matrix.

Our Feynman rules are: for every internal line we insert a
modified Feynman propagator ${\bar\Delta}_F(k)$ and integrate
over $k$ with the appropriate numerical factor.
We associate with every diagram a vertex
factor $V(p_1,...,p_n; k_1,...,k_n)$ where the $ps$ and $ks$
denote the external and internal momenta of the diagram,
respectively.

\section{\bf One Loop Self-energy Contribution}

The basic part of the one loop diagram for the
non-anticommutative scalar field theory with the potential
(\ref{potential}) is given by
\begin{equation}
\label{Sigma}
\Sigma=\frac{\lambda}{2(2\pi)^4}\int
d^4k\frac{i\exp[\frac{1}{2}(k\tau k)]}
{k^2-m^2+i\epsilon}.
\end{equation}
 
We now perform an analytic continuation in the complex $k$ and $p$ planes
with $k^2=-q^2=-(q_4^2+q_1^2+q_2^2+q_3^2)=-\kappa^2$ where $q$ denotes the Euclidean
momentum vector. Moreover, we have
$p^2=-\rho^2=-(p_4^{'2}+p_1^{'2}+p_2^{'2}+p_3^{'2})$. We treat $\tau^{\mu\nu}$ as a
non-degenerate `metric' with ${\rm det}(\tau^{\mu\nu})\not=0$, so that we can find an
orthonormal basis $v_1,...,v_n$ of the tangent space at each point such that
$\tau(v_\mu,v_\nu)=0$ if $\mu\not=\nu$ and $\tau(v_\mu,v_\nu)=\pm C$, where $C$ is a
constant. We choose the Lorentzian signature $\tau^{\mu\nu}=C\eta^{\mu\nu}$ and set
$\frac{1}{2}(k\tau k)\equiv
\frac{1}{2}k_\mu\tau^{\mu\nu}k_\nu=-\frac{\kappa^2}{\Lambda^2}$ and
$\frac{1}{2}(k\tau p)\equiv
\frac{1}{2}k_\mu\tau^{\mu\nu}p_\nu=-\frac{\kappa\rho}{\Lambda^2}$, where $\Lambda$ is
a constant with the dimensions of energy.
Since the numerator of ${\bar\Delta}_F(k)$
is an entire function, there are no additional singularities in the finite
momentum plane that can cause problems for the analytic continuation and
unitarity.

We get in Euclidean momentum space
\begin{equation}
\Sigma=\frac{\lambda}{16\pi^2}\int_0^\infty
d\kappa\frac{\kappa^3\exp\biggl(-\frac{\kappa^2}
{\Lambda^2}\biggr)}{\kappa^2+m^2} $$ $$
=\frac{\lambda}{32\pi^2}
\biggl[\Lambda^2-m^2\exp\biggl(\frac{m^2}{\Lambda^2}\biggr)
E_i\biggl(\frac{m^2}{\Lambda^2}\biggr)\biggr],
\end{equation}
where $E_i(x)$ is the exponential integral
\begin{equation}
E_i(x)=\int_1^\infty dt\frac{\exp(-xt)}{t}.
\end{equation}

The self-energy contribution $\Sigma$ can be expanded for
$\Lambda\gg m$ as
\begin{equation}
\Sigma=\frac{\lambda}{32\pi^2}\biggl\{\Lambda^2
-m^2\biggl[\gamma+\ln\biggl(\frac{m^2}{\Lambda^2}\biggr)\biggr]
+O\biggl(\frac{m^2}{\Lambda^2}\biggr)\biggr\}+O(\lambda^2).
\end{equation}

When we include the part with the vertex factor, we get
\begin{equation}
{\tilde\Sigma}(\rho)=
\frac{\lambda}{16\pi^2}\int_0^\infty d\kappa
\frac{\kappa^3\exp\biggl[-\biggl(\frac{\kappa^2}{\Lambda^2}
+\frac{3\kappa\rho}{2\Lambda^2}\biggr)\biggr]}{\kappa^2+m^2}.
\end{equation}
This integral is also convergent and yields a
finite self-energy contribution.

The vertex corrections in the non-anticommutative
$\lambda\phi_\diamondsuit^4$ theory will take the form
\begin{equation}
\Gamma(s)=\frac{\lambda}{2(2\pi)^4}\int
\frac{d^4k\exp\{\frac{1}{2}[(k\tau k)
+(l\tau l)]\}
V(p_1,p_2,p_3,p_4)}{(l^2-m^2+i\epsilon)(k^2-m^2+i\epsilon)},
\end{equation}
where $l=k-p$ and
\begin{equation}
s\equiv p^2=(p_1+p_2)^2,\quad t=(p_1-p_2)^2,\quad u=(p_1-p_4)^2
\end{equation}
are the Mandelstam variables. Then, we have for
the vertex correction contributions $\Gamma_1=\Gamma(s),
\Gamma_2=\Gamma(t),\Gamma_3=\Gamma(u)$. When we transform this
result into Euclidean momentum space, we find that the Gaussian
exponential factor in the internal momentum $k$ yields
convergent vertex corrections.

We see that in the
non-anticommutative scalar field theory, the one loop self-energy
diagram is {\it finite} for a generic value of the external
momentum $p$ and for fixed finite values of the parameter
$\Lambda$. The same holds true for the vertex one loop
corrections. When $\Lambda\rightarrow\infty$, we obtain the standard
classical spacetime continuum theory and the one loop self-energy is
quadratically divergent.

The convergence of both planar and non-planar loop
diagrams should hold to all orders of perturbation theory. Thus,
the self-energy corrections to the mass and the coupling constant
denoted by $\delta m^2$ and $\delta\lambda$ will be finite and
the `renormalization' of the mass and the coupling constant
will be a finite procedure.

An analysis of the unitarity condition for the non-anticommutative amplitudes shows
that it can be satisfied, because the propagator ${\bar\Delta}_F$ is modified only
by an {\it entire} function compared to the standard, commutative
propagator~\cite{Moffat4}. Thus, the position of singularities in the finite
region of the complex momentum plane are unaltered and the Cutkosky cutting rules can
be satisfied. The amplitudes avoid essential singularities at infinity and the
crossing symmetry relations are only physical at infinite energies, if we choose
$\tau^{mn}\not= 0$ and $\tau^{00}=\tau^{0n}=0$ $(m,n=1,2,3)$. We can then choose an
orthonormal frame such that $\tau^{mn}=-\delta^{mn}/\Lambda^2$, and the asymptotic
behavior of the modified propagator as $p^2\rightarrow\pm\infty$ is
\begin{equation}
{\bar\Delta}_F\sim \exp\biggl(-\frac{1}{2}{\bf p}^2/\Lambda^2\biggr),
\end{equation}
where ${\bf p}$ denotes the 3-momentum vector and ${\bf p}^2>0$. This choice of
orthonormal frame avoids the possibility of essential singularities in amplitudes at
infinite energies. However, it breaks Lorentz invariance but the Lorentz symmetry can
be broken `softly' by adding a Higgs spontaneous symmetry breaking contribution to
the action~\cite{Moffat5}.

\section{\bf Conclusions}

Classical spacetime geometry has been generalized to a superspace
with coordinates that are, in general, both noncommutative and
non-anticommutative, corresponding to their possessing `bosonic'
and `fermionic' degrees of freedom. The Moyal $\star$-product has
been extended to a $\circ$-product of field functions, defined in
terms of a nonsymmetric Hermitian tensor $\omega^{\mu\nu}$,
instead of the antisymmetric tensor $\theta^{\mu\nu}$ used in the
Moyal $\star$-product. We studied the simpler geometry in which
the antisymmetric tensor $\theta^{\mu\nu}=0$, and only the
symmetric tensor $\tau^{\mu\nu}$ contributes to a
$\diamondsuit$-product of field operators. The free-field theory
is different from the corresponding one in commutative field
theory, resulting in a modified Feynman propagator
${\bar\Delta}_F(k)$ and a modified vertex function
$V(k_1,...,k_n)$.

We restricted ourselves to the case $\tau^{mn}\not= 0$ and
$\tau^{00}=\tau^{0n}=0$, because only then do we get a field
theory that avoids potential problems with unitarity and
causality. The same circumstances exist for the noncommutative
field theories with the restriction that $\theta^{mn}\not= 0$ and
$\theta^{0n}=0$.

By calculating the one loop self-energy graph in
$\lambda\phi_\diamondsuit^4$
theory, we found that it is {\it finite} for both planar and
non-planar graphs. This result should hold for all loop graphs
due to the convergence of the loop diagrams, caused by the
exponential Gaussian factor associated with the modified
Feynman propagator.

The form of our modified Feynman propagator
${\bar\Delta}_F(k)$ involves an {\it entire} function of $k^2$. This means that
the theory is nonlocal, although the severity of the nonlocality has been
reduced due to the choice $\tau^{mn}\not= 0$ and
$\tau^{00}=\tau^{0n}=0$. A modified Feynman propagator of the
form \begin{equation}
i{\bar\Delta}_F(k)=\frac{i\exp\biggl(\frac{k^2-m^2}{\Lambda^2}\biggr)}
{k^2-m^2+i\epsilon}
\end{equation}
was used in the development of
a nonlocal field theory, which was perturbatively finite and for
Yang-Mills theory was gauge invariant to all
orders~\cite{Moffat3,Evans,Woodard}. However, this field theory
lacked a motivation for such a modified propagator, whereas in
the present scheme our quantization of spacetime, based on
non-anticommuting coordinates in a superspace, provides a
fundamental origin for the modified propagator and the
convergence of loop integrals. Nonlocal field theories of this
kind were studied extensively in the 1960s and 1970s
~\cite{Fainberg} and have also been the subject of investigations
in superstring theory and D-brane theory~\cite{Kapustin}.

In view of the finiteness of our loop diagrams, we do not find
any peculiar ultraviolet-infrared combination of divergences, as
is found in the noncommutative field theories~\cite{Susskind}.

The introduction of noncommutative and non-anticommutative coordinates
implies a natural unit of length $\ell$ in our superspace
formalism. The introduction of this unit of length makes it possible to
remove divergences in ordinary point particle, local field theory. To
accomplish this, we have given up the idea of a continuous spacetime
continuum, treating the coordinates of spacetime points as operators in a
supervector space. The introduction of a smallest unit of length in
spacetime, forces us to rescind the usual assumption of commutativity (or
anticommutativity) of coordinates $x^\mu$, otherwise the assumption of
Lorentz invariance of the {\it spectra} of the operators ${\hat x}^\mu$, if
they commute, implies continuous spectra corresponding to the measurements
of spacetime points.

It will be interesting to extend our investigation to gauge field theories
such as Yang-Mills theory and quantum gravity. By treating
quantum gravity as an effective perturbative
theory~\cite{Moffat2}, it is anticipated that due to the finite
results for the loop graphs obtained in scalar field theory,
there exists a strong possibility that a self-consistent,
effective quantum gravity theory can be derived,
based on a non-anticommutative spacetime geometry. This will be
the subject of a future paper.
\vskip 0.2 true in {\bf
Acknowledgments} \vskip 0.2 true in This work was supported by
the Natural Sciences and Engineering Research Council of Canada. I thank Craig
Burrell and Alex Williamson for helpful discussions.

\vskip 0.5 true in

\end{document}